\documentclass[journal]{IEEEtran}

\usepackage{graphicx}

\usepackage{cite}
\usepackage{amsmath}
\usepackage{url}
\usepackage{amsfonts}

\usepackage{subcaption}

\usepackage{multirow}
\usepackage{makecell}
\usepackage{booktabs}

\usepackage[rgb]{xcolor}
\usepackage{hyperref}
\hypersetup{
    colorlinks=true,
    linkcolor=blue,
    filecolor=magenta,
    urlcolor=cyan,
    citecolor=blue,
}
\usepackage{colortbl}

\def\X{{\mathbf{X}}}
\def\H{{\mathbf{H}}}
\def\Z{{\mathbf{Z}}}
\def\s{{\mathbf{s}}}

\def\Y{{\mathbf{Y}}}
\def\Rec{{\text{Recog}}}
\def\Syn{{\text{Synth}}}
\def\SpkEnc{{\text{SpkEnc}}}
\def\Dtrg{{\mathbf{D}_\text{trg}}}
\def\DVC{{\mathbf{D}_\text{VC}}}
\def\NPQ{{N_\text{PQ}}}

\begin{document}
\title{A Comparative Study of Self-supervised Speech Representation Based Voice Conversion}
\author{
        Wen-Chin~Huang,~\IEEEmembership{Member,~IEEE,}
        Shu-Wen~Yang,
        Tomoki~Hayashi,
        and~Tomoki~Toda% <-this % stops a space
\thanks{W.-C. Huang is with the Graduate School of Informatics, Nagoya University, Japan. E-mail: wen.chinhuang@g.sp.m.is.nagoya-u.ac.jp}% <-this % stops a space
\thanks{S.-W. Yang is with National Taiwan University, Taiwan.}% <-this % stops a space
\thanks{T. Tomoki and T. Toda are with Nagoya University, Japan.}% <-this % stops a space
% \thanks{X. Li is with Carnegie Mellon University, Pittsburgh PA, USA.}% <-this % stops a space
% \thanks{Manuscript received XXX XX, 2022; revised XXX XX, 2022.}
}

\markboth{Journal of \LaTeX\ Class Files,~Vol.~14, No.~8, August~2015}%
{Shell \MakeLowercase{\textit{et al.}}: Bare Demo of IEEEtran.cls for IEEE Journals}

\maketitle

\begin{abstract}
We present a large-scale comparative study of self-supervised speech representation (S3R)-based voice conversion (VC). In the context of recognition-synthesis VC, S3Rs are attractive owing to their potential to replace expensive supervised representations such as phonetic posteriorgrams (PPGs), which are commonly adopted by state-of-the-art VC systems. Using S3PRL-VC, an open-source VC software we previously developed, we provide a series of in-depth objective and subjective analyses under three VC settings: intra-/cross-lingual any-to-one (A2O) and any-to-any (A2A) VC, using the voice conversion challenge 2020 (VCC2020) dataset. We investigated S3R-based VC in various aspects, including model type, multilinguality, and supervision. We also studied the effect of a post-discretization process with k-means clustering and showed how it improves in the A2A setting. Finally, the comparison with state-of-the-art VC systems demonstrates the competitiveness of S3R-based VC and also sheds light on the possible improving directions.
% We believe the extensive analysis contributes to not only the S3R community but also the VC community.
\end{abstract}


\section{Introduction}

\IEEEPARstart{V}{oice} conversion (VC) refers to a technique that converts one type of speech to another while preserving the underlying spoken contents \cite{vc-survey, vc-survey-2021}. 
VC has a wide variety of applications, including accent conversion \cite{accent-conversion}, personalized speech synthesis \cite{personalized-tts, personalized-expresive-tts}, and speaking-aid device support \cite{EL-GMM, EL-o2m, EL-hybrid}.
In this work, we focus on the most widely investigated application of VC: speaker conversion, which refers to converting speech from a source speaker to a target speaker \cite{first-speaker-conversion}.

\begin{figure}[t]
	\centering
	\includegraphics[width=\linewidth]{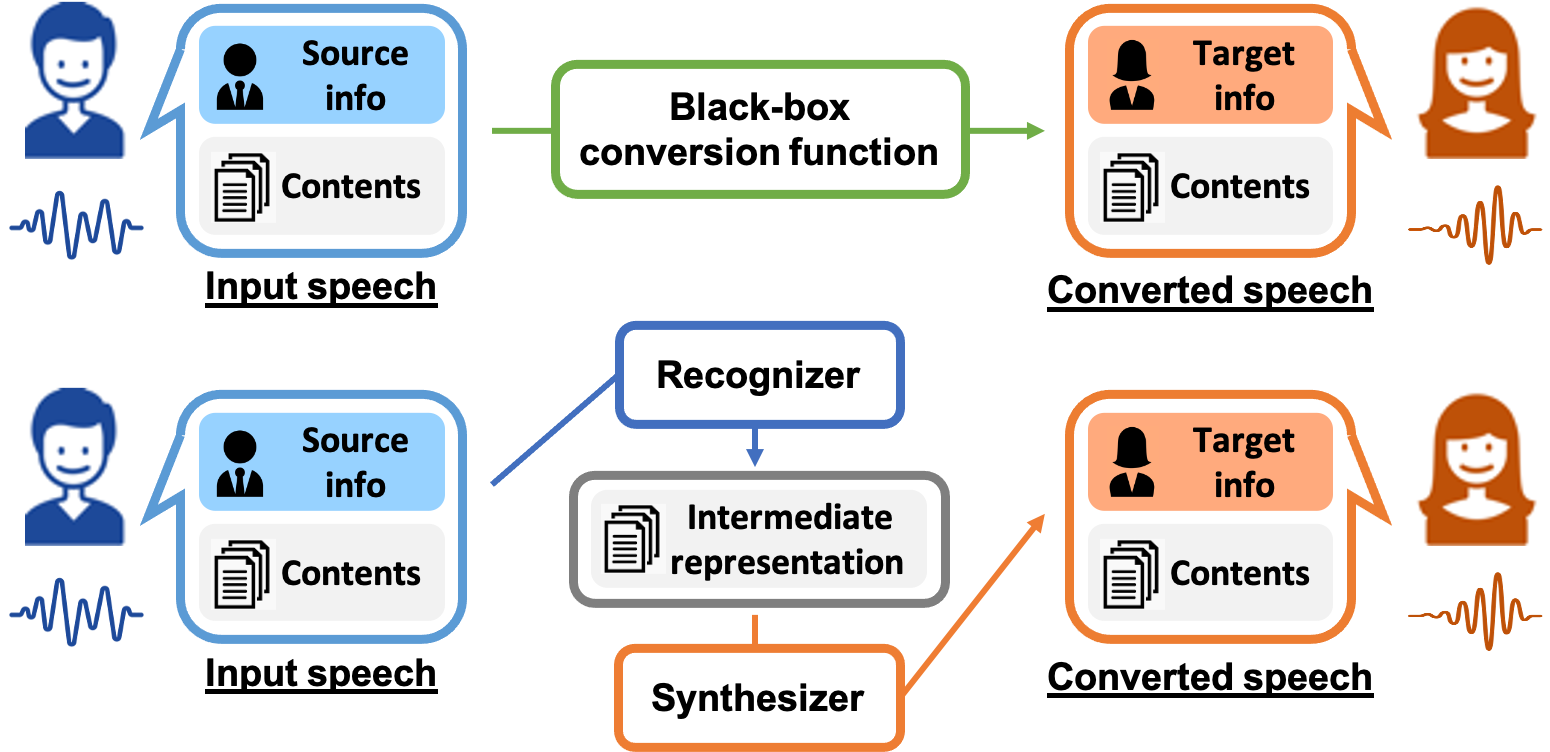}
	\caption{Top: black-box based voice conversion; bottom: decomposition by content disentanglement based voice conversion. \label{fig:blackbox-vs-rec-syn}}
\end{figure}

A widely-studied approach to VC aims at constructing a black-box function that directly maps source features into those of the target, as depicted in the top of Figure~\ref{fig:blackbox-vs-rec-syn}. Early studies employed statistical models such as Gaussian mixture models (GMMs) to represent such a function \cite{VC, GMM-VC}. To train the model, an alignment process with dynamic time warping must be performed beforehand \cite{vc-vq}, which requires access to a parallel training set containing utterances of the same linguistic contents from both source and target. To avoid the costly parallel data collection process, CycleGAN-based VC \cite{cyclegan-vc} was proposed to find the mapping function without explicit alignment using adversarial learning.

In recent years, a different strategy that has been gaining attention is to decompose the conversion function by disentangling the spoken contents from the others factors in speech, as depicted in the bottom of Figure~\ref{fig:blackbox-vs-rec-syn}. This is a reflection of the definition of VC: from the information perspective, VC can be performed by first extracting the spoken contents from the source speech, and then synthesizing the converted speech from the extracted contents with the characteristics of the target. Formally, starting from the source speech $\mathbf{X}$, a recognizer (or encoder) first extracts the spoken contents, $\mathbf{H}$, which is then consumed by the synthesizer (or decoder) to generate the converted speech, $\mathbf{Y}$:
\begin{equation}
    \Y = \Syn(\H), \H=\Rec(\X). \label{eq:formulation}
\end{equation}
Methods that implement this paradigm can be categorized based on how the two modules are optimized. For instance, a line of work tries to optimize the recognizer and synthesizer \textit{simultaneously} by using an autoencoding objective. In this framework, the ability of the encoder to extract linguistic contents are ensured by employing various information bottleneck, including variational autoencoder \cite{VAE, VAE-VC, VAE-GAN-VC}, vector quantization \cite{vqvae, VQVAE-code2spec} and instance normalization \cite{CHOU-M2MVC}.

In contrast, many have proposed to optimize the two modules \textit{separately}, and such an approach is often referred to as recognition-synthesis (rec-syn) based VC\footnote{To our konwledge, the term \textit{recognition-synthesis} was first defined in \cite{S2S-NP-VC}, which was only referred to VC systems composed by an ASR model and a speaker-dependent synthesizer. In this work, we define it to be any VC system that separately trains the recognizer and synthesizer.}. For instance, in the latest voice conversion challenge 2020 (VCC2020) \cite{vcc2020}, a baseline system and several top performing systems implemented such a framework \cite{vcc2020-asr-tts, vcc2020-task1-top, vcc2020-task2-top, vcc2020-srcb, vcc2020-casia}. It was shown in the challenge results that systems based on rec-syc VC were superior to autoencoder-based methods that trains the two modules concurrently in terms of both naturalness and similarity. Since these systems employed automatic speech recognition (ASR) models as the recognizer module, it is believed that the text data and the ASR objective function form a stronger information bottleneck for preserving the linguistic contents than the constraints used in the autoencoding framework.

% In rec-syn based VC, an ASR model trained on a labeled dataset is often used to extract the \textit{supervised} spoken content representation, such as text \cite{vcc2020-asr-tts} or phonetic posteriorgram (PPG) \cite{VC-PPG}.

One disadvantage of using ASR models as the recognizer module is the expensive dataset collection process. In low-resource settings such as the cross-lingual VC scenario \cite{vcc2020}, labeled datasets can be especially hard to collect. Therefore, researchers have resorted to unsupervised or the so-called self-supervised speech representation (S3R) learning paradigm, where a large-scale unlabeled data is used to learn rich, compact speech representations.

In addition to its label-free property, S3R based VC is also attractive in it being a good probing task for S3R analysis. Based on the information perspective of VC presented above, we may hypothesize that a good representation $\H$ in Eq.~\ref{eq:formulation} should be rich in content but contains little to none speaker information. As a result, an S3R model that can extract all-purpose speech representations may not be an optimal choice for VC. For instance, a well-known S3R, wav2vec 2.0 \cite{wav2vec2}, has been shown to be powerful in not only ASR but also speaker and language recognition \cite{wav2vec2-sid-lid}, implying that it encodes rich content, speaker and language information. Under our hypothesis, it may not be the best representation for VC. Such analyses may help researchers reach a better understanding of different S3R models.
% A recently published SUPERB benchmark \cite{superb} dedicates to compare different S3Rs across a range of \textit{discriminative} speech processing tasks, while it remains unclear what representations are optimal for \textit{generation} tasks like VC.

% In this paper, we describe S3PRL-VC, an extension of the S3PRL toolkit and SUPERB.
In this paper, we present a comperative study of S3R-based VC.
Our experiments were conducted using S3PRL-VC \cite{s3prl-vc}, an open-source VC software\footnote{\url{https://github.com/s3prl/s3prl/tree/master/s3prl/downstream/a2o-vc-vcc2020}} we previously developed that extended the SUPERB benchmark and the S3PRL toolkit \cite{superb}.
We conducted a large-scale evaluation, both objectively and subjectively, to analyze S3R-based VC systems from various aspects, including:
\begin{itemize}
    \item \textbf{Task}: Experiments were conducted under three kinds of settings: intra-/cross-lingual any-to-one (A2O) VC, where the system converts from an unseen speaker to a seen speaker of the same/different language, and intra-lingual any-to-any (A2A) VC, where both the source and target speakers are unknown during training. We used the VCC2020 dataset to unify the dataset condition, and to provide comparison with top systems in the challenge.
    % where the synthesizer is trained in a target-speaker-dependent fashion. We used the VCC2020 dataset, which allows us to test intra-lingual and cross-lingual settings. 
    % We also provide an any-to-any (A2A) extension by using an off-the-shelf d-vector \cite{d-vector} model to encode the unseen speaker information.
    \item \textbf{Model type}: We implemented models used in the top systems in VCC2018 \cite{VC-WNV-adapt} and VCC2020 \cite{vcc2020-task2-top}, which allows us to compare with the top systems in the respective years.
    \item \textbf{Multilinguality}: We validatethe cross-lingual transfer ability of S3Rs using the cross-lingual VC task. Furthermore, using the wav2vec 2.0 model, we compared the performance when trained on a mono-lingual and a multi-lingual dataset.
    \item \textbf{Supervision}: We provided results of supervised representations based systems using the same tasks and models, so we can understand the impact of supervision in recognizer training.
    \item \textbf{Discretization}: Although continuous features were used as default in the SUPERB benchmark, our initial investigation showed that they do not provide the sufficient disentanglement needed in the A2A setting. We then investigated a k-means based discretization used in \cite{speech-resynthesis}, and provided a comprehensive ablation study.
\end{itemize}
% Through subjective evaluations, it was shown that our A2O system was comparable to the top systems in VCC2020, and our A2A system was on par with a state-of-the-art A2A VC system \cite{s2vc}.
% S3PRL-VC is a competitive system by yielding (1) a comparable performance with VCC2020 top systems in the A2O setting in terms of similarity, and (2) state-of-the-art performance in S3R-based A2A VC.
% We believe that the findings in this paper can provide valuable insights to not only the VC community but also the S3R community.

This work aims to contribute to not only the VC field but also the S3R field. The contributions to the respective fields are summarized as follows:
\begin{itemize}
	\item \textbf{VC}: We aim at a unified, comprehensive study of S3R-based VC. Although getting increasingly popular in the VC field in recent years \cite{vqw2v-vc, speech-resynthesis, fragmentvc, s2vc, prosody-asr-tts}, each paper used their own experimental setting, including different datasets, models and evaluation protocol. As a result, it is difficult to compare different techniques to further identify drawbacks of current methods. Through this work, we hope to shed lights on a holistic understanding of the S3R-based VC framework, and provide a stepping stone for future VC researchers.
	\item \textbf{S3R}: We find VC suitable for investigating the disentanglement ability of S3R models. Most downstream tasks test one ability of the S3R model at a time, either the capability to encode rich and compact local content information (speech recognition, keyword spotting, etc.) or the power to represent global characteristics (speaker verification, emotion recognition, etc.) As stated above, we suspect VC can test these two abilities at once. Moreover, although we focus on speaker conversion in this work, by changing a task setting, it is possible to inspect the ability of the S3R model to disentangle different global attributes, such as accent or speaking style.
\end{itemize}

\section{Background and related works}

\begin{figure}[t]
	\centering
	\includegraphics[width=\linewidth]{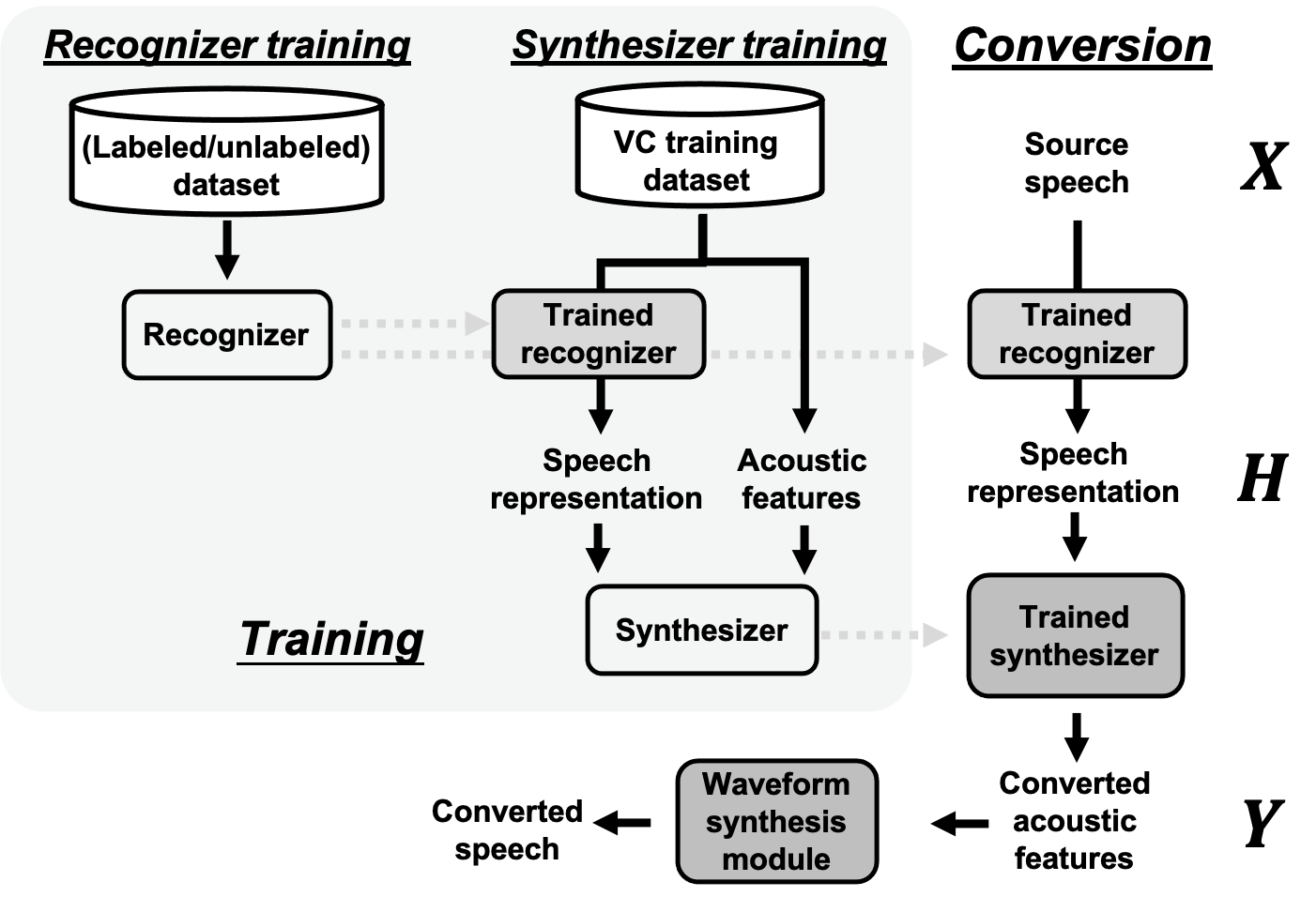}
	\caption{The training and conversion procedures of recognition-synthesis based VC. \label{fig:framework}}
\end{figure}

\subsection{Recognition-synthesis based voice conversion}
\label{ssec:rec-syn-vc}

Figure~\ref{fig:framework} illustrates the training and conversion processes in rec-syn based VC. The recognizer is first trained on a multi-speaker dataset, which can be either labeled or unlabeled. A common practice is to perform training in a speaker-independent fashion, which ensures the model's ability to encode the speech representation, $\H$, from any unseen speaker. Using the VC training dataset, $\DVC$, the synthesizer is trained to reconstruct the acoustic features from $\H$. Depending on the setting, the VC training dataset can be either a small target speaker dataset or a multi-speaker dataset, which we will describe later. In the conversion phase, the converted features, $\Y$, are generated following Eq.~\ref{eq:formulation}. The recognizer takes the source speech as input and extracts the S3Rs, which is consumed by the synthesizer to generate the converted acoustic features. Finally, a waveform synthesizer (ex. neural vocoder) generates the converted waveform.

In the literature, many types of intermediate representations have been used as $\H$, all of which have their own respective pros and cons. Table~\ref{tab:reprs} presents a comparison of the features based on various aspects. In the following we introduce three widely-used categories.

\subsubsection{Text}

Text is a straight-forward choice, as one can simply concatenate a pretrained ASR and text-to-speech (TTS) model. In VCC2020, one of the baseline systems called ASR+TTS \cite{vcc2020-asr-tts} and the top system of the intra-lingual task \cite{vcc2020-task1-top} both adopted text as the intermediate representation, and achieved outstanding performance in terms of similarity. This is mainly owing to the discrete and token-level nature of text. Since prosodic information including the speaking rate and the pitch pattern are discarded after recognition, the synthesizer needs to use a powerful model like sequence-to-sequence (seq2seq) to reconstruct the target characteristics. However, this approach suffers from mispronunciation when the accuracy of the ASR and TTS model is insufficient, as shown in \cite{vcc2020-asr-tts}. There are also VC scenarios where the source style needs to be preserved, such as singing VC \cite{prosody-transfer-vc}.

\subsubsection{Phonetic posteriorgrams or bottleneck features}

Phonetic posteriorgrams (PPGs) were first applied to VC in \cite{VC-PPG}. PPGs represent the frame-wise posterior probabilities of each phonetic class, which are derived from the acoustic model (AM) of an HMM based ASR model. The training target of the AM are phoneme labels, so only the output of the last layer of the AM has the physical meaning of PPG, but some have proposed to use the ouptut from other layers. For example, the system in \cite{S2S-iFLYTEK-VC} used the output before the softmax layer and referred to them as bottleneck features (BNFs). Either PPGs or BNFs are frame-level continuous features, thus better perserve the linguistic contents and can help produce high-quality speech. The top system in VCC2018 \cite{VC-WNV-adapt} and the top system in VCC2020 task 2 \cite{vcc2020-task2-top} both adopted this feature.
% The ability to retain more prosodic informations makes it possibl to apply to emotional VC.
However, the frame-level nature makes the conversion of speaking rate difficult. Efforts needed for the frame-level labels of the ASR dataset also raised the difficulty of constructing the system. 

\subsubsection{Self-supervised speech representations}
\label{sssec:s3r-vc}

To reduce the labeling cost of training ASR models, applying S3Rs to VC has become increasing popular. Being free from labeled data not only reduces the labeling cost, but also makes it possible to use more unlabeled datasets and work under low-resource settings. S3Rs have been applied to a wide verity of VC settings, including any-to-one VC \cite{vqw2v-vc}, many-to-many VC \cite{speech-resynthesis}, any-to-any VC \cite{fragmentvc, s2vc} and cross-lingual VC \cite{prosody-asr-tts}.

The typical usage of S3R models is to extract continuous features for downstream tasks. However, due to the lack of supervision, continuous S3Rs lack the ability to fully separate contents from other factors such as speaker identity, resulting in poor performance in the A2A setting \cite{s3prl-vc}. One way to provide the sufficient disentanglement is through discretization, as shown in \cite{VQVC}. Certain S3R models such as VQVAE \cite{vqvae} or vq-wav2vec\cite{vq-wav2vec} are able to generate discrete outputs due to their architecture, thus some have therefore proposed VC systems based on them \cite{vqw2v-vc, vcc2020-as}. However, not all S3R models have such discretization design. Recently, \cite{speech-resynthesis} proposed to apply a k-means based post-discretization process on the continuous S3Rs. The learned discrete units were shown to be effective in a many-to-many VC setting.

\begin{table}[t]
	\centering
	\caption{A comparison of intermediate representations in recognition-synthesis based voice conversion.}
	
	\centering
	\begin{tabular}{| c | c | c | c |}
		\hline
		Representation & Text & \makecell{Phonetic\\Posteriorgram} & \makecell{Self-supervised\\speech representations} \\
		\hline
		Extractor & \multicolumn{2}{c|}{ASR model} & self-supervised model \\
		\hline
		Training data & \multicolumn{2}{c|}{labeled data} & unlabeled data \\
		\hline
		Resolution & token level & \multicolumn{2}{c|}{frame level} \\
		\hline
		Continuous? & discrete & continuous & can be either \\
		\hline
		Examples & \cite{vcc2020-asr-tts, vcc2020-task1-top} & \cite{vcc2020-task2-top, vcc2020-casia, vcc2020-srcb} & \cite{speech-resynthesis, vqw2v-vc, fragmentvc, s2vc, prosody-asr-tts, vcc2020-as} \\
		\hline
	\end{tabular}
	\label{tab:reprs}
\end{table}

\subsection{Self-supervised speech representation learning}

In recent years, self-supervised learning has been the state-of-the-art approach in various research fields. It implies a principle that first pretrains an \textit{upstream} model that learns general knowledge by solving self-supervised tasks on a large amount of unlabeled data, followed by fine-tuning prediction layers on various \textit{downstream} tasks\footnote{In the context of S3R-based VC, the recognizer is represented and the synthesizer are represented by the S3R upstream model and the downstream prediction layers, respectively. In the remainder of this paper, we will use these two terms interchangeably.}. When applied to speech, S3Rs are expected to capture linguistic, speaker, prosodic, and semantic information of speech. In the literature, though with different network architectures, S3Rs are commonly grouped by their objective functions. Generative modeling incorporates language model-like training losses to predict unseen regions (such as future or masked frames), in order to maximize the likelihood of the observed data. Examples include APC \cite{apc}, VQ-APC \cite{vq_apc}, Mockingjay \cite{mockingjay}, TERA \cite{tera}, and NPC \cite{npc}. Discriminative modeling aims to discriminate (or contrast) the target unseen frame with randomly sampled ones, which is equivalent to mutual information maximization. Examples include CPC \cite{CPC, modified-cpc}, wav2vec \cite{wav2vec}, vq-wav2vec \cite{vq-wav2vec}, wav2vec 2.0 \cite{wav2vec2} and HuBERT \cite{hubert}. Finally, multi-task learning applies multiple objectives, including waveform generation, prosody features regression and contrastive InfoMax. PASE+ \cite{pase+} is the most representative approach. 

\section{Tasks Design}

\subsection{General description of VCC2020}
\label{ssec:vcc2020}

\begin{table}[t]
	\centering
	\caption{Summary of the data conditions in VCC2020.}
	
	\centering
	\begin{tabular}{ c | c c | c c }
		\toprule
		\multirow{2}{*}[-2pt]{Task} & \multicolumn{2}{c|}{Training phase} & \multicolumn{2}{c}{Conversion phase} \\
		\cmidrule(lr){2-5}
		& Source & Target & Source & Converted \\
		\midrule
		Task 1 & \multirow{2}{*}[-8pt]{\makecell{70 Eng.\\utterances}} & \makecell{70 Eng.\\utterances} & \multirow{2}{*}[-8pt]{\makecell{25 Eng.\\utterances}} & \multirow{2}{*}[-8pt]{\makecell{25 Eng.\\utterances}} \\
		\cmidrule(lr){1-1} \cmidrule(lr){3-3}
		Task 2 & & \makecell{70 Man./Ger./Fin.\\utterances} & & \\
		\bottomrule
	\end{tabular}
	\label{tab:data-vcc2020}
\end{table}

% \begin{table}[t]
% 	\centering
% 	\caption{Summary of the data conditions in VCC2020.}
	
% 	\centering
% 	\hspace*{-0.3cm}
% 	\begin{tabular}{| c | c | c | c | c |}
% 		\hline
% 		\multirow{2}{*}[-2pt]{Task} & \multicolumn{2}{c|}{Training phase} & \multicolumn{2}{c|}{Conversion phase} \\
% 		\cline{2-5}
% 		& Source & Target & Source & Converted \\
% 		\hline
% 		Task 1 & \multirow{2}{*}[-6pt]{\makecell{70 Eng.\\utterances}} & \makecell{70 Eng.\\utterances} & \multirow{2}{*}[-6pt]{\makecell{25 Eng.\\utterances}} & \multirow{2}{*}[-6pt]{\makecell{25 Eng.\\utterances}} \\
% 		\cline{1-1} \cline{3-3}
% 		Task 2 & & \makecell{70 Man./Ger./Fin.\\utterances} & & \\
% 		\hline
% 	\end{tabular}
% 	\label{tab:data-vcc2020}
% \end{table}

All experiments in this work are benchmarked on the VCC2020 dataset \cite{vcc2020}. There are two tasks in VCC2020, with intra-lingual VC being task 1 and cross-lingual VC being task 2.
The data conditions are summarized in Table~\ref{tab:data-vcc2020}.
The two tasks share the same two English male and female source speakers. The target speakers include two male and two female English speakers for task 1, and one male and one female speaker each of Finnish, German, and Mandarin for task 2.
For each speaker, 70 utterances (roughly five minutes) in their respective languages and contents are provided, and there are 25 test sentences for evaluation.
During conversion, the source speech (which is in English) is converted as if it was uttered by the target speaker while keeping the linguistic contents unchanged.

\subsection{Intra-lingual and cross-lingual any-to-one VC}

We first consider the two tasks in VCC2020 under the A2O setting. A2O VC aims to convert from any arbitrary speech into that of a predefined target speaker. 
As mentioned in \ref{ssec:rec-syn-vc}, the ability to encode $\H$ from any unseen speaker is ensured by the common practice of training S3Rs on a multi-speaker dataset.
In the A2O setting, the VC training dataset in Figure~\ref{fig:framework} is the target speaker dataset, $\Dtrg$. The synthesizer is trained to reconstruct the acoustic feature from $\H$. As described in Secion~\ref{ssec:vcc2020}, the language of $\Dtrg$ is English and Finnish/German/Mandarin in the intra-lingual and cross-lingual setting, respectively.
% In the conversion phase, the converted features, $\Y$, are generated following Eq.~\ref{eq:formulation}.
% the recognizer takes the source speech as input and extracts the S3Rs, which is consumed by the synthesizer to generate the converted acoustic features.
% Finally, a waveform synthesizer (ex. neural vocoder) generates the converted waveform.

A2O VC is a good probing task to investigate several characteristics of an upstream S3R model. A fundamental requirement of VC is the linguistic consistency, so there is a positive correlation between the VC performance of an S3R model and its ability to faithfully encode $\H$.
% Second, if an S3R model encodes rich speaker information, then the source speaker information in $\X$ will conflict with the target speaker attributes injected by the synthesizer, which hurts the VC performance.
Also, during the synthesizer training in cross-lingual VC, the S3R model may fail to generalize to $\X$ from a non-English target speaker since most existing S3R models are trained with English datasets only. It is worthwhile to examine the ability of mono-lingual S3R models to transfer to different languages.

% \begin{figure*}[t]
%     \begin{minipage}[t]{0.5\textwidth}
%         \centering
%     	\includegraphics[width=\columnwidth]{imgs/post-discretization-all.png} 
%     	\centering
%     	\caption{Left: the post-discretization process overview. Top right: the cluster ensemble technique. Bottom right: the product quantization techniques.}
%     	\label{fig:post-discretization}
% 	\end{minipage}
% 	~
% 	\begin{minipage}[t]{0.5\textwidth}
%     	\centering
%     	\includegraphics[width=\columnwidth]{imgs/all_models_dvec} 
%     	\caption{The models implemented in this work. Left: the simple model. Middle: the simple model with an AR loop. Right: the Tacotron2 model, with extension to an any-to-any model by accepting a d-vector as the speaker embedding. \label{fig:models}}
% 	\end{minipage}
% \end{figure*}

\subsection{Intra-lingual any-to-any VC}

We then extend the VCC2020 dataset for an A2A scenario, also known as zero-shot VC. A2A VC attempts to convert to a target speaker where $\Dtrg$ is so limited (less than one minute) such that fine-tuning in infeasible.
In this setting, the $\DVC$ used to train the A2A VC model is a separate multi-speaker dataset.
As in the A2O setting, the synthesizer is trained to reconstruct the acoustic feature from $\H$. However, due to the speaker-independent nature of S3Rs, $\H$\ does not provide sufficient speaker information to recover the speaker information. Thus, the input is augmented with a speaker embedding, $\s$, extracted by an off-the-shelf speaker encoder, which is pretrained on an automatic speaker verification (ASV) dataset and objective.
% Such a paradigm is also used in zero-shot TTS \cite{adaptation-verification}.
In training, the speaker embedding extracted from the target waveform is used. During conversion, given $\Dtrg$, $\s$ is formed as an average of each embedding from each utterance. We may then rewrite Eq.~\ref{eq:formulation} as:
\begin{equation}
    \vspace{-0.1cm}
    \Y = \Syn(\H, \s), \H=\Rec(\X), \s=\SpkEnc(\Dtrg). \label{eq:a2a-formulation}
\end{equation}

A2A VC helps us investigate how complete can an S3R model filter out the speaker information, which is an important ability in rec-syn based VC. We explain why the A2O setting cannot explore this ability well.
% since it performs \textit{model-based} conversion.
Imagine the scenario where a lot of speaker information remains in the S3R. Since the training target is always the target speaker dataset, it is possible that the model removes the speaker information first then inject back to the output. However, in the A2A VC scenario, the training target is drawn randomly from the multi-speaker dataset, thus a ``speaker-free'' S3R is more demanding. That is to say, during conversion, if an S3R model encodes rich speaker information, then the source speaker information in $\X$ will conflict with the target speaker attributes injected by the synthesizer, which hurts the VC performance.

\section{Implementations}

\subsection{Recognizers (upstream models)}

Table~\ref{tab:obj} depicts the list of S3Rs we compared in this work, which are the upstream models supported in S3PRL at the date of publication. For a complete list of information (training data, architecture, objective, etc.), refer to \cite{superb}.
All upstreams are trained with English data, mostly LibriSpeech \cite{librispeech} or LibriLight \cite{librilight}.
In addition to the S3Rs, two extra upstreams were included: (1) mel-spectrogram, ``mel'', and (2) ``PPG (TIMIT)'', which is trained supervisedly on the TIMIT dataset \cite{timit}.

\subsection{Synthesizer model implementation}
\label{ssec:synthesizer}

\begin{figure}[t]
	\centering
	\includegraphics[width=\columnwidth]{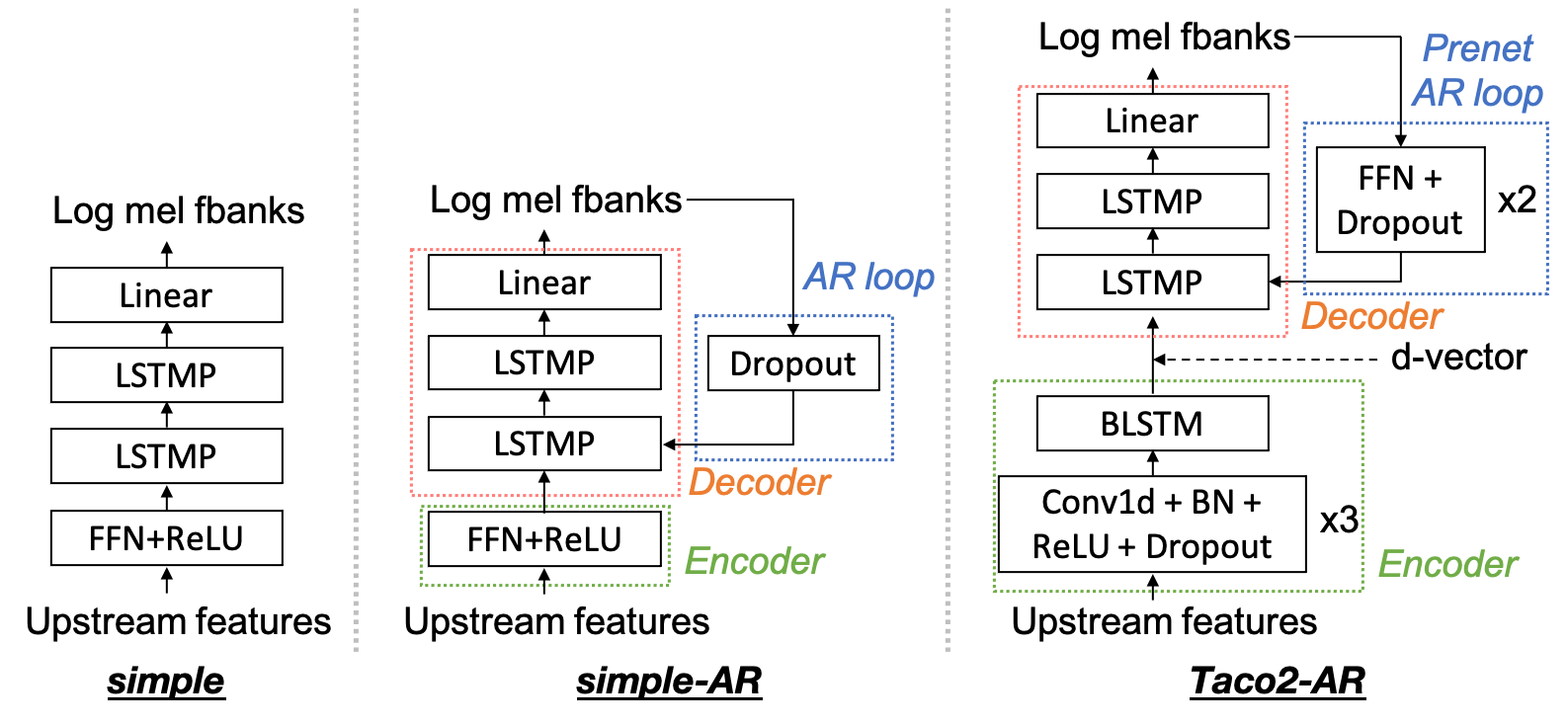} 
	\caption{The models implemented in this work. Left: the simple model. Middle: the simple model with an AR loop. Right: the Tacotron2 model, with extension to an any-to-any model by accepting a d-vector as the speaker embedding. \label{fig:models}}
% 	\vspace{-0.5cm}
\end{figure}

Log mel fbanks was selected as the target acoustic feature. We implemented several models to resemble top systems of past VCCs, as illustrated in Figure~\ref{fig:models}. We avoid expensive model components like attention \cite{transformer} because (1) fast benchmarking is a key requirement of SUPERB/S3PRL, and (2) the frame-level feature used in this framework frees us from changing the temporal structure. For discrete inputs generated by the methods described in Section~\ref{ssec:post-discretization}, they are embedded using lookup tables first.
\begin{itemize}
	\item \textbf{Simple}: We start from the model used by the top system in VCC2018 \cite{VC-WNV-adapt}. The simple model consists of a single layer feed-forward network (FFN), two long short-term memory layers with projection (LSTMP), and a linear projection layer. 
	\item \textbf{Simple-AR}: As autoregressive (AR) modeling has been shown to be effective in speech synthesis \cite{ar-rnn-mdn-spss}, we added an AR loop to the simple model. At each time step, the previous output is consumed by the first LSTMP layer. Dropout is essential in the AR loop to avoid exposure bias brought by teacher-forcing \cite{ar-f0-spss, Taco}.
	\item \textbf{Taco2-AR}: We increase the model complexity by using a model architecture similar to that of Tacotron 2 \cite{Taco2}, which resembles the model used by the top system in VCC2020 \cite{vcc2020-task2-top}. Different from Tacotron 2, the attention module was not used as it was reported to be useless in \cite{vcc2020-task2-top}.
\end{itemize}

\subsection{Post-discretization process for any-to-any VC}
\label{ssec:post-discretization}

\begin{figure}[t]
	\centering
	\includegraphics[width=\columnwidth]{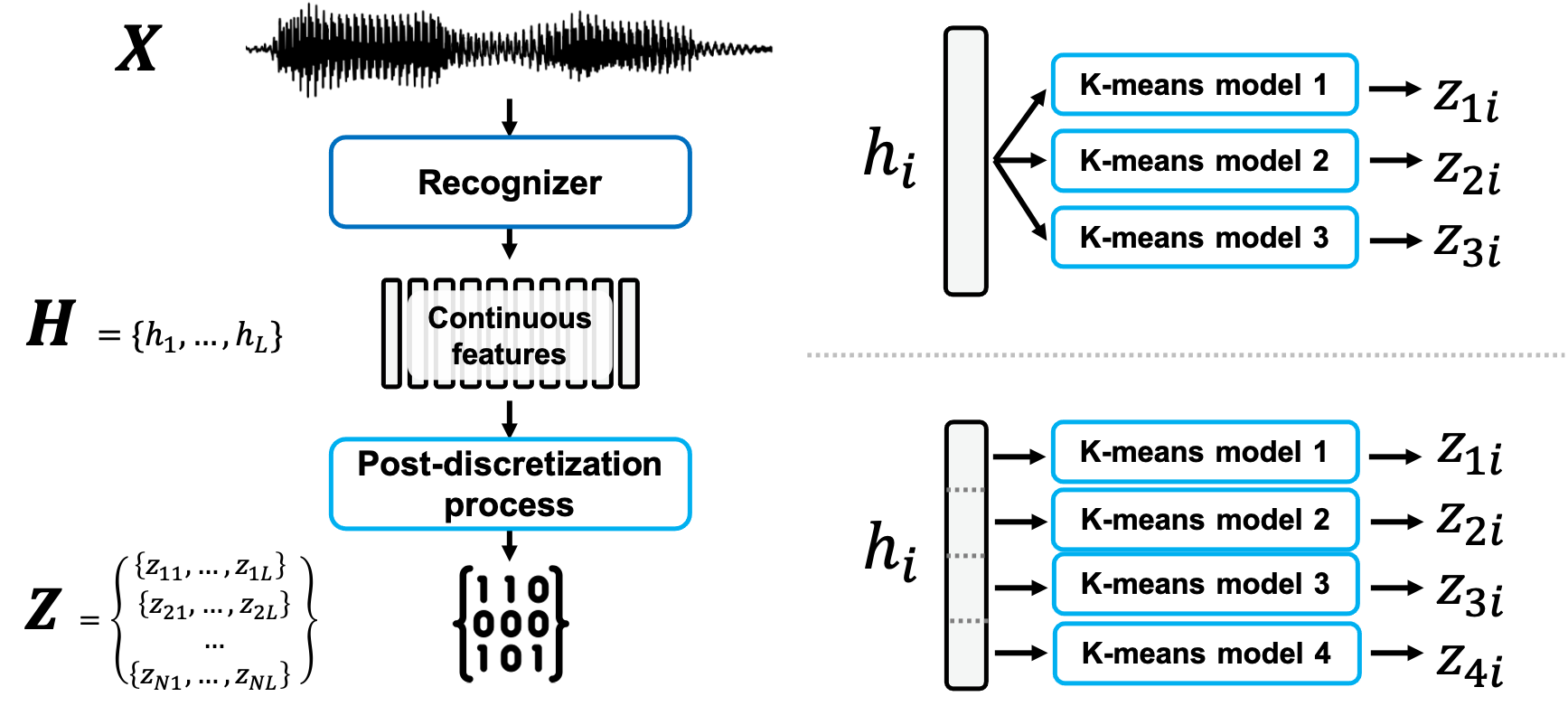} 
	\caption{Left: the post-discretization process overview. Top right: the cluster ensemble technique with 3 k-means models using different numbers of clusters. Bottom right: the product quantization techniques with 4 partitions.}
    \label{fig:post-discretization}
\end{figure}

In our initial investigations \cite{s3prl-vc}, using continuous features cannot satisfy the disentanglement requirement in the A2A scenario. As a result, most S3Rs fail to convert the speaker identity, as we show in later sections. We thus provide an extension to match the tendency in the A2A setting to that in A2O.

We impose a stronger information bottleneck by adopting the post-discretization process proposed in \cite{speech-resynthesis}. Specifically, as illustrated in the left of Figure~\ref{fig:post-discretization}, the k-means clustering algorithm takes the continous features $\H$\ returned by the recognizer, and returns corresponding discrete indices $\Z$\ using a codebook of size $K$ trained with a separate dataset in advance.

However, in our preliminary experiments, the method proposed in \cite{speech-resynthesis} performs poorly when applied to certain S3Rs. The generated speech often suffers from poor the intelligibility, even when using a large codebook. We suspect that the information bottleneck introduced by discretization is too strong. To offer more expressive power, inspired by \cite{hubert}, we employ the following two additional techniques shown in the right of Figure~\ref{fig:post-discretization}.
Both methods try to describe one feature vector with multiple k-means models (i.e. multiple indices) to increase the degree of freedom. In the experimental section, we provide a complete investigation of these two techniques.

\subsubsection{Cluster ensemble}

Using an ensemble of k-means models with different codebook sizes can capture different granularity, and each k-means model can provide complementary information to back each other up. Specifically, given a continuous feature vector $h_i$, we use $N_{\text{CE}}$ k-means models to generate $N_\text{CE}$ indices: $[z_{1i}, z_{2i}, \dots, z_{N_\text{CE}i}]$, where the codebook of $n$-th model has size $K_n$ clusters. Each $K_n$ should be set to different numbers, so that different k-means models can learn to capture different levels of detail.

\subsubsection{Product quantization}

Product quantization (PQ) is a technique where the feature space is partitioned into multiple subspaces, and each subspace is quantized separately using different k-means models. Specifically, given a continuous feature vector $h_i\in\mathbb{R}^{d}$, we first partition it into $N_\text{PQ}$ subvectors: $[h_{1i},h_{2i},\dots,h_{N_\text{PQ}i}]$ where each subvector has size $h_{ni}\in\mathbb{R}^{d/N_\text{PQ}}$. Then, each subvector is consumed by a separate k-means model to generate $N_\text{PQ}$ indices: $[z_{1i}, z_{2i}, \dots, z_{N_\text{PQ}i}]$. The k-means models can be of different numbers of clusters as done in cluster ensemble, but for simplicity, here we set all k-means models to have equal number of clusters.

\subsection{Other implementation setups}

\subsubsection{Any-to-any VC settings}

The dataset used to train the A2A VC model is the VCTK dataset \cite{vctk}. For the speaker encoder, we used the d-vector model \cite{d-vector} trained on a mix of datasets, including LibriSpeech, VoxCeleb 1 \cite{voxceleb1} and 2 \cite{voxceleb2}. For the post-discretization process, following \cite{speech-resynthesis}, all k-means models are trained on the LibriSpeech clean-100h set \cite{librispeech}. Although some studies use intermediate layer outputs for discretization \cite{speech-resynthesis, wav2vec-u}, for simplicity, we use the last outputs for all S3R models.

\subsubsection{Waveform synthesizer}

We used the HiFi-GAN \cite{hifigan}, a state-of-the-art parallel real-time neural vocoder. For the A2O setup, we mixed the data of all 14 speakers in VCC2020 with the VCTK dataset, while for the A2A setup we used only the VCTK dataset.

\begin{table*}[ht!]
\centering

%\tiny
%\scriptsize
\footnotesize
%\small

\caption{Objective evaluation results on different VC settings over various S3Rs using \textbf{continuous} features. For MCD and WER, the smaller the better; for ASV, the higher the better.}
\label{tab:obj}
\hspace*{-9pt}
\begin{tabular}{|l||r|r|r|r|r|r|r|r|r|r|r|r|r|r|}
\hline
\multirow{3}{*}{Upstream}  & \multicolumn{9}{c|}{Intra-lingual A2O} & \multicolumn{2}{c|}{Cross-lingual A2O} & \multicolumn{3}{c|}{Intra-lingual A2A} \\ \cline{2-15}
& \multicolumn{3}{c|}{Simple} & \multicolumn{3}{c|}{Simple-AR} & \multicolumn{3}{c|}{Taco2-AR} & \multicolumn{2}{c|}{Taco2-AR} & \multicolumn{3}{c|}{Taco2-AR} \\ \cline{2-15}
& MCD  & WER  & ASV  & MCD  & WER  & ASV  & MCD  & WER  & ASV  & WER & ASV & MCD  & WER  & ASV  \\ \hline \hline
%& MCD $\downarrow$ & WER $\downarrow$ & ASV $\uparrow$ & MCD $\downarrow$ & WER $\downarrow$ & ASV $\uparrow$ & MCD $\downarrow$ & WER $\downarrow$ & ASV $\uparrow$ & MCD $\downarrow$ & WER $\downarrow$ & ASV $\uparrow$ & MCD $\downarrow$ & WER $\downarrow$ & ASV $\uparrow$ \\ \hline \hline

mel & 8.41 & 48.5 & 59.00 & 8.92 & 22.7 & 49.75 & 8.47 & 38.3 & 77.25 & 39.0 & 46.67 & 9.49 & 4.2 & 19.50 \\
PPG (TIMIT) & 7.78 & 69.0 & 85.50 & 7.83 & 58.9 & 95.25 & 7.18 & 33.6 & 99.75 & 51.0 & 84.67 & \textbf{8.31} & 12.9 & \textbf{83.50} \\ \hline
PASE+ \cite{pase+} & 9.29 & 5.0 & 26.75 & 9.52 & 5.7 & 26.00 & 8.66 & 30.6 & 63.20 & 36.3 & 34.67 & 9.85 & 4.2 & 8.00 \\
APC \cite{apc} & 8.67 & 8.6 & 48.00 & 8.73 & 7.1 & 41.75 & 8.05 & 27.2 & 87.25 & 33.9 & 52.33 & 9.57 & 3.5 & 23.25 \\
VQ-APC \cite{vq_apc} & 8.12 & 10.8 & 81.25 & 8.37 & 7.4 & 60.50 & 7.84 & 22.4 & 94.25 & 28.4 & 68.00 & 9.43 & 4.0 & 22.00 \\
NPC \cite{apc} & 7.74 & 39.0 & 92.75 & 8.15 & 21.1 & 76.75 & 7.86 & 30.4 & 94.75 & 37.6 & 59.00 & 9.39 & 4.4 & 21.00 \\
Mockingjay \cite{mockingjay} & 8.58 & 31.3 & 51.00 & 8.74 & 9.5 & 47.00 & 8.29 & 35.1 & 79.75 & 39.2 & 46.00 & 9.43 & 5.0 & 25.00 \\
TERA \cite{tera} & 8.60 & 11.4 & 46.50 & 8.67 & 6.0 & 42.50 & 8.21 & 25.1 & 83.75 & 29.2 & 49.33 & 9.31 & 5.2 & 18.75 \\
Modified CPC \cite{modified-cpc} & 8.71 & 9.4 & 40.00 & 8.87 & 7.0 & 30.00 & 8.41 & 26.2 & 71.00 & 35.3 & 32.83 & 9.61 & 4.1 & 10.75 \\
DeCoAR 2.0 \cite{decoar2} & 8.31 & 7.4 & 54.75 & 8.33 & 6.4 & 53.00 & 7.83 & 17.1 & 90.75 & 26.8 & 59.33 & 9.28 & 4.0 & 27.00 \\
wav2vec \cite{wav2vec} & 7.45 & 14.0 & \textbf{95.50} & 7.64 & 4.9 & 90.50 & 7.45 & 10.1 & 98.25 & 13.9 & 75.83 & 8.77 & 3.5 & 40.00 \\
vq-wav2vec \cite{vq-wav2vec} & \textbf{7.41} & 13.4 & 91.00 & \textbf{7.24} & 11.6 & \textbf{98.75} & \textbf{7.08} & 13.4 & \textbf{100.00} & 21.0 & \textbf{88.83} & 8.47 & 4.2 & 73.25  \\
wav2vec 2.0 Base \cite{wav2vec2}& 7.80 & 24.7 & 92.75 & 7.77 & 5.0 & 86.50 & 7.50 & 10.5 & 98.00 & 14.9 & 82.17 & 9.03 & 3.2 & 27.00 \\
wav2vec 2.0 Large & 7.64 & 12.5 & 81.75 & 7.67 & 9.0 & 82.75 & 7.63 & 15.8 & 97.25 & 22.7 & 78.00 & 8.99 & 4.1 & 22.25 \\
% XLSR & 9.85 & 104.0 & 0.50 & 7.63 & 12.5 & 75.50 & 7.52 & 14.1 & 97.75 & 24.2 & 72.50 & 9.01 & 3.7 & 27.25  \\
HuBERT Base \cite{hubert} & 7.70 & \textbf{5.5} & 89.25 & 7.79 & \textbf{4.7} & 84.25 & 7.47 & \textbf{8.0} & 98.50 & \textbf{13.5} & 82.33 &  9.19 & 3.4 & 23.25 \\
HuBERT Large & 7.54 & 5.6 & 95.00 & 7.54 & 5.6 & 93.00 & 7.22 & 9.0 & 99.25 & 15.9 & 86.50 & 9.13 & \textbf{3.0} & 27.75 \\
\hline
\end{tabular}
% \vspace{-0.5cm}
\end{table*}

% \section{Main experimental evaluation results}
\section{Experimental evaluation results}

In this section, we first describe the evaluation metrics (Section~\ref{ssec:metrics}). Then we provide a series of complete objective evaluations and a large-scale listening test to analyze \textit{continuous} feature-based S3R-based VC and to compare with state-of-the-art systems (Section~\ref{ssec:compare-sota}). The aspects we investigate include the synthesizer model type (Section~\ref{ssec:compare-models}), multilinguality (Section~\ref{ssec:compare-multilinguality}) and supervision (Section~\ref{ssec:compare-supervision}). We finally examine the effectiveness of the post-discretization process (Sections~\ref{ssec:effect-discretization} and~\ref{ssec:continuous-vs-discrete}).

\subsection{Evaluation metrics and protocols}
\label{ssec:metrics}

\subsubsection{Objective evaluation}

We employed the following three objective evaluation metrics, all of which measure different aspects of a VC system. For the cross-lingual A2O task, we did not report the MCD results.
\begin{itemize}
    \item \textbf{MCD}: The mel cepstrum distortion (MCD) is an intrusive, L2-norm based metric based on the mel cepstrum coefficient (mcep) which measures the general performance:
	\begin{equation}
	    \text{MCD} [dB] = \frac{10}{\log 10} \sqrt{2 \sum_{d=1}^K (\text{mcep}^{(c)}_d - \text{mcep}^{(t)}_d )^2} ,
	\end{equation}
	where $K$ is the dimension of the mceps and $\text{mcep}^{(c)}_d$ and $\text{mcep}^{(t)}_d$ represent the $d$-th dimensional coefficient of the converted mceps and the target mceps, respectively. The WORLD vocoder \cite{WORLD} was used to extract the mceps.
% 	A dynamic time warping (DTW) based alignment is performed to find the corresponding frame pairs between the non-silent converted and target MCC sequences, and the utterance-wise MCD is then calculated.
	\item \textbf{WER}: The word error rate (WER) is a non-intrusive measure of the intelligibility and the linguistic consistency of the converted speech. We used a pretrained wav2vec 2.0 model\footnote{Performance and APIs can be found at \url{https://huggingface.co/facebook/wav2vec2-large-960h-lv60-self}}.
	\item \textbf{ASV}: The accept rate from a pretrained ASV model measures whether the speaker identity is converted by calculating the cosine similarity using speaker embeddings \cite{vcc2020-prediction}. Specifically, the cosine similarity of the d-vectors extracted from each converted utterance and the corresponding reference are calculated. We then report the percentage of the testing utterances whose cosine similarity exceeds a pre-calculated threshold.
\end{itemize}

\subsubsection{Subjective evaluation}

For the subjective test, we asked listening participants to evaluate two common aspects in VC: naturalness and similarity.
Listeners were asked to evaluate the naturalness on a five-point scale.
For conversion similarity, a natural target speech and a converted speech were presented, and listeners were asked to judge whether the two samples were produced by the same speaker on a four-point scale.
For each system, a total of 80 utterances (5 random $\times$ 16 conversion pairs)  were evaluated. Recordings of the target speakers were also included in the naturalness test and served as the upper bound. 
We used an open-source toolkit \cite{p808-open-source} that implemented the ITU-T Recommendation P.808 \cite{p808} to screen unreliable ratings obtained through the Amazon Mechanical Turk (Mturk). We recruited more than 280 listeners from the United States and had each sample rated by five different participants on average.
Audio samples are available online\footnote{\url{https://unilight.github.io/Publication-Demos/publications/s3prl-vc/}}.

\subsection{Comparison of different synthesizer model types}
\label{ssec:compare-models}

We first investigate the impact of using different synthesizer models described in Section~\ref{ssec:synthesizer} in the intra-lingual A2O setting, as shown in Table~\ref{tab:obj}.
First, only by adding the AR loop to the Simple model, most S3Rs benefit from large improvements in WER. With Taco2-AR, all S3Rs except PASE+ and modified CPC achieved an ASV accept rate higher 80\%, while all S3Rs suffered from a degradation in WER.
This shows that increasing the model capacity can significantly improve the speaker similarity, while sacrificing the intelligibility.
However, we would like to emphasize that: (1) the WER is a strict measurement of intelligibility, and human can actually recognize better than machine. After listening to the samples, our internal percepion was that compared to simple-AR, the quality was greatly improved and intelligibility degradation was not as serious as shown in the table.
(2) the Taco2-AR model yields the best MCD scores, which, as we will show later, correlates better with subjective naturalness and similarity.
% For me, after listening to the samples, I feel the quality was greatly improved and intelligibility degradation was actually not that bad.
% Nonetheless, by comparing the simple model and the taco2-AR model, all S3Rs except NPC enjoyed a large improvement in MCD.
(3) we empirically found the training time of the three models similar.
Based on these reasons, we decided to use the Taco2-AR model for the succeeding tasks and comparisons.

\subsection{Investigation on model multilinguality}
\label{ssec:compare-multilinguality}

\begin{table}[t]
\centering
\caption{Comparison of wav2vec 2.0 trained on mono-lingual data and cross-lingual data in the cross-lingual A2O scenario, using the Taco2-AR model. The results of wav2vec 2.0 Large are extracted from Table~\ref{tab:obj}.}
\label{tab:mono-vs-cross}

\begin{tabular}{|l|l||r|r|}
\hline
Upstream & Training data size & WER & ASV \\

\hline \hline
% wav2vec 2.0 Base & LibriSpeech 960 hr & 14.9 & 82.17 \\
wav2vec 2.0 Large & LibriLight 60k hr & 22.7 & 78.00 \\
XLSR \cite{xlsr} & 56k hr from 53 languages & 24.2 & 72.50 \\ 
\hline
\end{tabular}
\end{table}

% Next, we compare the results of using S3Rs for different tasks. We first compare cross-lingual VC with the intra-lingual setting.
Next, we assess the VC performance of S3R models in the cross-lingual setting.
Looking again at Table~\ref{tab:obj}, we first find S3Rs trained on a mono-lingual corpus can still work well in the cross-lingual setting, demonstrating their ability to transfer across languages.
However, compared with the intra-lingual A2O task, it could be clearly observed that all S3Rs degraded in terms of both the WER and ASV accept rate in the cross-lingual setting.
In VCC2020, it was also reported that cross-lingual VC is indeed a harder task than intra-lingual VC, as the listening test results of all participating teams were much worse.

To further investigate the impact of the training data language, in Table~\ref{tab:mono-vs-cross} we report the results of XLSR \cite{xlsr}, a model that has the same architecture as wav2vec 2.0 Large but trained on a mixture of datasets from 53 language, resulting in 56k hours of data.
We found that compared to wav2vec 2.0 Large trained on mono-lingual data, XLSR was not particularly good. We suspect that when the training set is large enough, the model can already capture the variations among all languages such that a multilingual dataset will not be needed. Also, since the source language during conversion is English, it is possible that monolingual models are sufficient. It is worthwhile investigating this point by considering a different setting in the future, such as converting from non-English languages.
% Another interesting finding from Table~\ref{tab:mono-vs-cross} (and also Table~\ref{tab:obj}) is that wav2vec 2.0 Large performs worse than wav2vec 2.0 Base. This is contrary to the tendency in ASR, as shown in the original paper \cite{wav2vec2}. A possible explanation is that the representation from a larger model trained with a larger, more diverse is more suitable for solving classification problems like ASR. This may also explain why XLSR did not outperform wav2vec 2.0 Large.

% Finally, in the intra-lingual A2A setting, all S3Rs yielded WERs much lower than those in the A2O setting, while the MCD values and ASV accept rates were significantly worse. Even the best upstream, vq-wav2vec, yielded only an accept rate of $73.25$. One possible reason is that in the A2A VC setting, modern S3Rs still fail to disentangle content, such that the synthesizer preserves too much speaker information. Another reason may be that a jointly trained speaker encoder \cite{s2vc} is essential for S3R-based VC.

\begin{table}[t]
\centering

%\tiny
\scriptsize
% \footnotesize
%\small

\caption{Comparison with state-of-the-art systems. All upstreams use the Taco2-AR model. Nat. and Sim. stand for naturalness and similarity, respectively. Both Nat. and Sim. are the higher the better. The objective results (MCD, WER, ASV) are extracted from Table~\ref{tab:obj}.}

\label{tab:comparison}
% \hspace*{-10pt}
\begin{tabular}{|>{\scriptsize}l||r|r|r|c|c|}
\hline
System & MCD & WER & ASV & Nat. & Sim. \\

\hline \hline
\multicolumn{6}{|c|}{Intra-lingual A2O} \\ \hline
mel & 8.47 & 38.3 & 77.25 & 2.61 $\pm$ .11 & 35$\%$ $\pm$ 3$\%$ \\ 
PPG (TIMIT)& 7.18 & 33.6 & 99.75 & 3.32 $\pm$ .10 & 58$\%$ $\pm$ 4$\%$ \\
\hline
PASE+ & 8.66 & 30.6 & 63.20 & 2.58 $\pm$ .12 & 31$\%$ $\pm$ 3$\%$ \\
APC & 8.05 & 27.2 & 87.25 & 2.92 $\pm$ .11 & 43$\%$ $\pm$ 4$\%$ \\
VQ-APC & 7.84 & 22.4 & 94.25 & 3.08 $\pm$ .10 & 40$\%$ $\pm$ 4$\%$ \\
NPC & 7.86 & 30.4 & 94.75 & 2.98 $\pm$ .11 & 46$\%$ $\pm$ 3$\%$ \\
Mockingjay & 8.29 & 35.1 & 79.75 & 2.81 $\pm$ .12 & 42$\%$ $\pm$ 4$\%$ \\
TERA & 8.21 & 25.1 & 83.75 & 2.91 $\pm$ .12 & 37$\%$ $\pm$ 4$\%$ \\
Modified CPC & 8.41 & 26.2  & 71.00  & 2.74 $\pm$ .11 & 33$\%$ $\pm$ 3$\%$ \\
DeCoAR 2.0 & 7.83  & 17.1  & 90.75  & 3.04 $\pm$ .11 & 43$\%$ $\pm$ 4$\%$ \\
wav2vec & 7.45 & 10.1 & 98.25 & 3.40 $\pm$ .05 & 52$\%$ $\pm$ 2$\%$ \\
vq-wav2vec & 7.08 & 13.4 & 100.00 & 3.59 $\pm$ .10 & 59$\%$ $\pm$ 4$\%$ \\
wav2vec 2.0 B. & 7.50 & 10.5 & 98.00 & 3.36 $\pm$ .06 & 51$\%$ $\pm$ 2$\%$ \\
wav2vec 2.0 L. & 7.63 & 15.8 & 97.25 & 3.26 $\pm$ .10 & 50$\%$ $\pm$ 4$\%$ \\
% XLSR & 7.52 & 14.1 & 97.75 & 3.40 $\pm$ 0.10 & 51$\%$ $\pm$ 4$\%$ \\
HuBERT B. & 7.47 & 8.0 & 98.50 & 3.48 $\pm$ .10 & 55$\%$ $\pm$ 4$\%$ \\
HuBERT L. & 7.22 & 9.0 & 99.25 & 3.47 $\pm$ .10 & 54$\%$ $\pm$ 4$\%$ \\
\hline
USTC-2018$\dagger$ \cite{VC-WNV-adapt} & -- & 6.5 & 99.00 & 4.20 $\pm$ .08 & 55$\%$ $\pm$ 4$\%$ \\
USTC-2020 \cite{vcc2020-task1-top} & 6.98 & 5.4 & 100.00 & 4.41 $\pm$ .07 & 82$\%$ $\pm$ 3$\%$ \\
SRCB \cite{vcc2020-srcb} & 8.90 & 11.5 & 92.00 & 4.16 $\pm$ .08 & 68$\%$ $\pm$ 3$\%$ \\
CASIA \cite{vcc2020-casia} & 7.13 & 11.0 & 98.25 & 4.25 $\pm$ .08 & 61$\%$ $\pm$ 4$\%$ \\
ASR+TTS \cite{vcc2020-asr-tts} & 6.48 & 8.2 & 100.00 & 3.84 $\pm$ .09 & 75$\%$ $\pm$ 3$\%$ \\
\hline
Target & -- & 0.7 & -- & 4.57 $\pm$ 0.14 & -- \\ 

\hline \hline
\multicolumn{6}{|c|}{Cross-lingual A2O} \\ \hline
PPG (TIMIT)& -- & 51.0 & 84.67 & 2.79 $\pm$ .08 & 43$\%$ $\pm$ 3$\%$ \\
\hline
vq-wav2vec & -- & 21.0 & 88.83 & 3.28 $\pm$ .08 & 44$\%$ $\pm$ 3$\%$ \\
HuBERT L. & -- & 15.9 & 86.50 & 3.13 $\pm$ .08 & 41$\%$ $\pm$ 3$\%$ \\
\hline
USTC-2018 \cite{VC-WNV-adapt} & -- & 5.6 & 97.67 & 4.17 $\pm$ .06 & 34$\%$ $\pm$ 3$\%$ \\
USTC-2020 \cite{vcc2020-task2-top} & -- & 7.6 & 96.00 & 4.27 $\pm$ .07 & 43$\%$ $\pm$ 3$\%$ \\
SRCB \cite{vcc2020-srcb} & -- & 8.6 & 78.67 & 4.34 $\pm$ .07 & 34$\%$ $\pm$ 3$\%$ \\
CASIA \cite{vcc2020-casia} & -- & 10.5 & 91.67 & 4.11 $\pm$ .07 & 45$\%$ $\pm$ 3$\%$ \\
ASR+TTS \cite{vcc2020-asr-tts} & -- & 34.5 & 67.83 & 2.51 $\pm$ .08 & 39$\%$ $\pm$ 3$\%$ \\
\hline
Target & -- & -- & -- & 4.48 $\pm$ 0.12 & -- \\

\hline \hline
\multicolumn{6}{|c|}{Intra-lingual A2A} \\ \hline
PPG (TIMIT) & 8.32 & 12.7 & 84.25 & 3.41 $\pm$ .08 & 34$\%$ $\pm$ 4$\%$ \\
\hline
vq-wav2vec & 8.47 & 4.2 & 73.25 & 3.58 $\pm$ .09 & 28$\%$ $\pm$ 3$\%$ \\
\hline
S2VC$\dagger$ \cite{s2vc} & -- & 12.4 & 71.50 & 2.90 $\pm$ .09 & 29$\%$ $\pm$ 3$\%$ \\
\hline
\multicolumn{6}{l}{\makecell[l]{$\dagger$: Systems generate 16kHz, so MCD is not calculable and direct score\\comparison should be made with caution.}}\\
\end{tabular}
% \vspace{-0.5cm}
\end{table}

\begin{table}[t]
\centering
\caption{Linear correlation coefficients between different metrics.}
\label{tab:correlation}

\begin{tabular}{|c||c|c|c|c|c|}
\hline
Metric & MCD & WER & ASV & Nat. & Sim. \\
\hline \hline
MCD & -- & 0.678 & -0.934 & -0.968 & -0.961 \\
WER & -- & -- & -0.640 & -0.808 & -0.587 \\
ASV & -- & -- & -- & 0.910 & 0.911 \\
Nat. & -- & -- & -- & -- & 0.932 \\
Sim. & -- & -- & -- & -- & -- \\
\hline
\end{tabular}
% \vspace{-0.5cm}
\end{table}

\subsection{Comparing with state-of-the-art systems using subjective evaluation}
\label{ssec:compare-sota}

We then compared S3R-based VC models with state-of-the-art systems. \textbf{USTC-2018} \cite{VC-WNV-adapt}, \textbf{USTC-2020} \cite{vcc2020-task1-top, vcc2020-task2-top}\footnote{USTC's systems used text and PPG for the intra-lingual and cross-lingual tasks, respectively.}, \textbf{SRCB} \cite{vcc2020-srcb}, \textbf{CASIA} \cite{vcc2020-casia} were top systems in VCC2020, all of which adopted PPGs, synthesizer pretraining on a multi-speaker dataset, and AR vocoders. Notably, they used thousands of hours of internal data for training. \textbf{ASR+TTS} \cite{vcc2020-asr-tts} was the seq2seq+non-AR vocoder baseline in VCC2020. \textbf{S2VC} \cite{s2vc} is the STOA system for A2A VC. The results are shown in Table~\ref{tab:comparison}. We summarize our observations as follows:
\begin{itemize}
    \item vq-wav2vec outperformed all other upstreams in the subjective test, with a 3.59 naturalness and 59$\%$ similarity in the intra-lingual A2O setting.
    \item In the A2O settings, there was still a naturalness gap between vq-wav2vec and other VCC2020 top systems (3.59 v.s. 4.16-4.25, 3.28 v.s. 4.11-4.34). As for similarity, vq-wav2vec was on par with USTC-2018 and CASIA in the intra-lingual A2O setting, and achieved top in the cross-lingual setting.
    \item In the A2A setting, vq-wav2vec was on par with S2VC in similarity, while being significantly better in naturalness. Our system is therefore the new state-of-the-art in S3R-based A2A VC.
\end{itemize}

% \vspace{-0.1cm}
\subsection{Impact of supervision}
\label{ssec:compare-supervision}

Although top systems using PPG greatly outperformed vq-wav2vec in naturalness, they used AR vocoders and the system was trained on large internal datasets, so the impact of supervision is not yet clear. To this end, we compared vq-wav2vec result with ``PPG (TIMIT)'' and the same vocoder. From Table~\ref{tab:comparison}, we first find ``PPG (TIMIT)'' has a high WER and a low naturalness score, shoing that it was indeed of low quality. Nonetheless, in all three settings, ``PPG (TIMIT)'' can achieve similar or higher similarity scores than vq-wav2vec. This shows that supervision greatly contributes to similarity, especially in a difficult setting like A2A VC.
This also shows that the ability of current S3Rs to disentangle speaker information is still limited when compared to PPG, and can be further improved in the future.
That being said, we can still achieve good performance without supervision if the S3R was designed properly. 
% We can also infer that the quality of the vocoder plays a key role to filling the gap.
% which are not publicly accessible to other researchers. 

\vspace{-0.1cm}
\subsection{Justify the objective metrics with correlation analysis}
\label{ssec:justify}

Conducting a subjective test whenever a new S3R is developed cannot meet the fast benchmark requirement of SUPERB. Therefore, we examine if the objective measures align well with human perception. Using the intra-lingual A2O results over different upstreams, we calculated pairwise linear correlation coefficients. Results in Table~\ref{tab:correlation} suggested that MCD best aligned with both naturalness and similarity.
Note that in this correlation analysis, we considered systems that used the same synthesizer and neural vocoder. Since the correlation result is strongly affected by the pool of methods evaluated in a listening test, this good correlation could be observed only in such a homogeneous condition. That is to say, as long as the synthesizer and the vocoder are the same, we can safely use the objective measures to compare different upstreams. This implication is very useful for the benchmarking requirement of SUPERB.

% \section{Ablation studies on the post-discretization process}

% \subsection{Effectiveness of cluster ensemble and product quantization}
\subsection{Investigation of the post-discretization process}
\label{ssec:effect-discretization}

\begin{table}[t]
\centering
\caption{Results of HuBERT Base and Mockingjay using the cluster ensemble and the product quantization techniques in the any-to-any scenario, with the Taco2-AR model. The best numbers within the same upstream are in bold face.}
\label{tab:cluster-ensemble-pq}

\begin{tabular}{|l|l|l||r|r|r|}
\hline
Upstream & \# clusters ($K_n$) & $N_\text{PQ}$ & MCD & WER & ASV \\
\hline \hline
\multirow{10}{*}[0pt]{HuBERT Base} & 50 & \multirow{7}{*}[0pt]{1} & 8.41 & 22.0 & 79.50 \\
& 100 & & 8.25 & 10.3 & 83.50 \\
& 200 & & 8.32 & 10.2 & 84.25 \\
& 50+100 & & 8.37 & 10.2 & \textbf{86.25} \\
& 50+200 & & 8.28  & 8.6  & 84.25 \\
& 100+200 & & 8.29 & 8.8 & 83.25 \\
& 50+100+200 & & 8.40 & 7.9 & 85.00 \\
\cline{2-6}
& 50 & \multirow{3}{*}[0pt]{2} & 8.37 & 12.7 & 84.50 \\
& 100 & & \textbf{8.23} & 8.2 & \textbf{86.25} \\
& 200 & & 8.32 & \textbf{7.2} & 81.75 \\
\hline
\multirow{5}{*}[0pt]{Mockingjay} & 100 & \multirow{3}{*}[0pt]{1} & 9.12 & 77.4 & 63.00 \\
& 200 & & 9.10 & 73.1 & \textbf{63.25} \\
& 50+100+200 & & 9.02 & 59.7 & 61.00 \\
\cline{2-6}
& 100 & \multirow{2}{*}[0pt]{2} & 9.07 & 64.5 & 59.00 \\
& 200 & & \textbf{8.95} & \textbf{55.4} & 61.75 \\
\hline
\end{tabular}
\end{table}

In Table~\ref{tab:cluster-ensemble-pq}, we report results of applying cluster ensemble and PQ on two upstreams, namely HuBERT Base and Mockingjay, in the A2A setting. First, we can observe that the intelligibility (WER) improves when the number of k-means model in the ensemble increases. That is to say, using two k-means models is better than using one, and using three is even better. The intelligibility is also improved when using PQ, and the improvement is consistent across all numbers of clusters. However, using more k-means models in both cluster ensemble and PQ means to loosen the speaker information bottleneck, which can harm the conversion similarity (ASV) as well as MCD. Finally, an interesting finding is that by only partitioning into two feature subvectors, the MCD and WER are still better than using an ensemble of three k-means models, suggesting that PQ is a more effective method then cluster ensemble. This is consistent with the finding in \cite{hubert}. We thus use PQ in the following experiments.

% \subsection{Impact of number of partitions}
% \label{ssec:npq}

\begin{table}[t]
\centering
\caption{Results of HuBERT Base and Mockingjay varying the number of partitions ($N_\text{PQ}$) in the product quantization technique. The number of clusters is set to 200 in all k-means models. The task is any-to-any VC, and the model is the Taco2-AR model.}
\label{tab:pq}

\begin{tabular}{|l|l||r|r|r|}
\hline
Upstream & $N_\text{PQ}$ & MCD & WER & ASV \\
\hline \hline
\multirow{9}{*}[0pt]{HuBERT Base} & 1& 8.32& 10.2& 84.25\\
& 2 & 8.32 & 7.2 & 81.75 \\
& 4 & 8.39 & 5.8 & 84.00 \\
& 8 & 8.35 & \textbf{3.5} & \textbf{84.50} \\
& 16 & \textbf{8.31} & 4.1 & 78.00 \\
& 32 & 8.41 & 3.6 & 75.00 \\
& 64 & 8.45 & 3.8 & 75.25 \\
& 128 & 8.38 & 3.9 & 74.00 \\
& 256 & 8.37 & 4.2 & 74.75 \\
\hline
\multirow{6}{*}[0pt]{Mockingjay} & 1 & 9.10 & 73.1 & \textbf{63.25} \\
& 2 & \textbf{8.95} & 55.4 & 61.75 \\
& 4 & 9.09 & 37.8 & 52.50 \\
& 8 & 9.14 & 20.2 & 39.25 \\
& 16 & 9.25 & 12.8 & 34.75 \\
& 32 & 9.37 & \textbf{8.6} & 29.75 \\
\hline
\end{tabular}
\end{table}

\begin{figure}[t]
	\centering
	\includegraphics[width=\columnwidth]{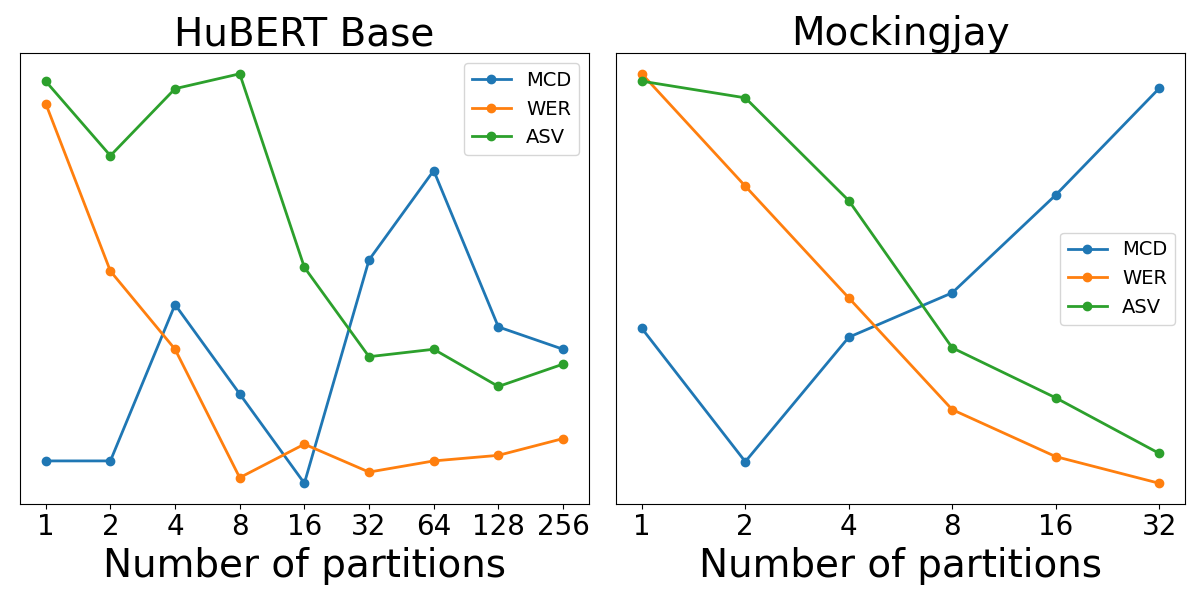} 
	\caption{Visualizing the effect of number of partitions. Left: HuBERT Base. Right: Mockingjay. \label{fig:pq}}
% 	\vspace{-0.5cm}
\end{figure}

Based on the observations in Table~\ref{tab:cluster-ensemble-pq}, we then investigate how much speaker information are leaked when the number of partitions increases. Table~\ref{tab:pq} shows the results varying the number of partitions using HuBERT and Mockingjay, and Figure~\ref{fig:pq} is a visualization of the overall trend. For HuBERT Base, we can first observe a diminishing returns effect in WER. That is to say, the WER stops to improve when $\NPQ$ is large enough. We can also observe that the conversion accuracy stays at a similar level when $\NPQ$ is small, and starts to drop when $\NPQ$ gets larger. These observtions show that we can find an optimal $\NPQ$ such that the WER is optimized while maintaining a similar level of conversion accuracy. However, for Mockingjay, both WER and ASV are monotonically decreasing, such that we cannot find such an optimal point by only looking at these two metrics. As a result, we resolve to MCD to find the optimal $\NPQ$.

\subsection{Comparison of continuous and discrete features}
\label{ssec:continuous-vs-discrete}

\begin{table}[t]
\centering

%\tiny
%\scriptsize
\footnotesize
%\small

\caption{Results on any-to-any VC with continuous and discrete features over various upstreams. The results using continuous features are extracted from Table~\ref{tab:obj}.}
\label{tab:continuous-vs-discrete}

\begin{tabular}{|l||r|r|r|r|r|r|}
\hline
\multirow{2}{*}{Upstream}  & \multicolumn{3}{c|}{Continuous} & \multicolumn{3}{c|}{Discrete} \\ \cline{2-7}
& MCD  & WER  & ASV  & MCD  & WER  & ASV \\ \hline \hline

% mel & 9.49 & 4.2 & 19.50 \\
% PPG (TIMIT) & \textbf{8.31} & 12.9 & \textbf{83.50} \\ \hline
PASE+ & 9.85 & 4.2 & 8.00 & 8.92 & 81.7 & 74.00 \\
APC & 9.57 & 3.5 & 23.25 & 8.66 & 22.4 & 81.25 \\
VQ-APC & 9.43 & 4.0 & 22.00 & 8.42 & 21.0 & 85.50 \\
NPC & 9.39 & 4.4 & 21.00 & 8.78 & 46.0 & 74.50 \\
Mockingjay & 9.43 & 5.0 & 25.00 & 8.95 & 55.4 & 61.75 \\
TERA & 9.31 & 5.2 & 18.75 & 8.40 & 37.1 & 67.00 \\
Modified CPC & 9.61 & 4.1 & 10.75 & 8.69 & 13.8 & 75.50 \\
DeCoAR 2.0 & 9.28 & 4.0 & 27.00 & -- & -- & --$\dagger$ \\
wav2vec & 8.77 & 3.5 & 40.00 & 8.34 & 15.2 & \textbf{86.50} \\
vq-wav2vec & \textbf{8.47} & 4.2 & \textbf{73.25} & 8.49 & 22.5 & 82.50 \\
wav2vec 2.0 B. & 9.03 & 3.2 & 27.00 & 8.90 & 54.3 & 75.75 \\
wav2vec 2.0 L. & 8.99 & 4.1 & 22.25 & 8.97 & 67.7 & 72.75 \\
% XLSR & 9.01 & 3.7 & 27.25  \\
HuBERT B. & 9.19 & 3.4 & 23.25 & 8.31 & \textbf{4.1} & 78.00\\
HuBERT L. & 9.13 & \textbf{3.0} & 27.75 & \textbf{8.23} & 7.4 & 86.25 \\
\hline
\multicolumn{7}{l}{\makecell[l]{$\dagger$: Fails to be trained.}}\\
\end{tabular}
\end{table}

Finally, we compare the results in the A2A setting when using continuous and discrete features. Since there are too many hyperparemeters that can be tuned, we applied the PQ technique and set the number of clusters to be 200, and we searched the best $\NPQ$ between 1, 2 and 4 with the lowest MCD. We report the results in Table~\ref{tab:continuous-vs-discrete}. It can be clearly observed that the post-discretization process indeed serves as a strong speaker information bottleneck as the ASV scores of all S3Rs are significantly higher than the continuous counterpart. As described in Section~\ref{ssec:post-discretization}, most S3Rs suffer from poor intelligibility even with the PQ technique. However, certain S3Rs still achieved an acceptable balance of intelligibility and conversion similarity, resulted in MCD values lower than that of the best performing continuous S3R (8.47 from vq-wav2vec), such as VQ-APC, wav2vec, HuBERT Base and HuBERT Large.

\section{Discussion and conclusion}

In the paper, we presented a comparative study of S3R-based VC. We used S3PRL-VC, an extension of the S3PRL toolkit that focused on the VC downstream task. We evaluated the S3Rs under the context of VC, and provided a series of in-depth analysis in various aspects including the synthesizer model type, different VC tasks, supervision and discretization. We also compared with the state-of-the-art VC systems in VCC2020, and showed that there is still room for improvement in terms of quality and similarity.

Readers from different research communities can gain individual insights from this work.
From the VC perspective, in S3PRL-VC, to meet the fast benchmarking requirement, some techniques that were shown to be effective were not applied, such as fine-tuning target speaker-dependent vocoders \cite{VC-WNV-adapt, si-wnv}, training the synthesizer with waveform domain losses \cite{joint-conversion-wnv, speech-resynthesis}, or fine-tuning the vocoder with ground truth aligned synthesis \cite{Taco2, refined-vae-wnv, hifigan}. That is to say, the performance can be further optimized. In addition, applications to other VC tasks such as emotional VC, expressive VC, singing VC and VC for speaking aid devices are also worth investigating.

From the S3R perspective, we have shown that there are certain challenges that are required by VC, such as the preservation of the spoken contents and the disentanglement of speaker information. It is therefore worthwhile to continue to use VC as a probing task when designing new S3R models. 

Finally, we would like to discuss the special position of VC in the context of the recent SUPERB \cite{superb-sg} activities. SUPERB is a collection of benchmark resources that aims to evaluate S3Rs across various speech tasks, with an assumption in mind that different representations should outperform others in different tasks due to their pretext-task nature. However, in the original version that consisted of only 10 discriminative tasks, it turned out that wav2vec 2.0 and HuBERT outperformed all other S3Rs. This dominance was broken after the introduction of VC, where vq-wav2vec was shown to be the best in the A2O setting, due to its disentangling ability.

This finding has several important implications. First, it shows that VC can be used to examine the disentanglement performance of a S3R, and there is a need for disentanglement if one tries to develop an universal representation, which not yet exists. Also, we hope this work can serve as a good initiative for future S3R researchers to emphasize on the disentanglement performance of their model, without hurting the scores on other tasks like ASR and ASV. This could have a bigger impact on the community compared to pursuing incremental improvements on other tasks.

\section*{Acknowledgment}

We would like to thank the S3PRL/SUPERB team for the fruitful discussions. This work was partly supported by JSPS KAKENHI Grant Number 21J20920, JST CREST Grant Number JPMJCR19A3, and a project, JPNP20006, commissioned by NEDO, Japan.

\ifCLASSOPTIONcaptionsoff
  \newpage
\fi



\bibliographystyle{IEEEtran}
\bibliography{ref}


% \end{IEEEbiography}


%\newpage



%\vfill


\end{document}